%Paper: hep-ph/9307351
%From: "Jean Cleymans" <CLEYMANS@physci.uct.ac.za>
%Date: 28 Jul 93 14:40:51 GMT+0200

\magnification=1200\overfullrule=0pt\baselineskip=12pt
\vsize=22truecm \hsize=15truecm \overfullrule=0pt\pageno=0

\font\titlefont=cmbx12 scaled \magstep1
\font\sectnfont=cmbx10  scaled \magstep1
\font\subsectnfont=cmbx8  scaled \magstep1
\font\eightrm=cmr8
\long\def\fussnote#1#2{{\baselineskip=9pt
     \setbox\strutbox=\hbox{\vrule height 7pt depth 2pt width 0pt}%
     \eightrm
\footnote{#1}{#2}}}
\def\mname{\ifcase\month\or January \or February \or March \or April
           \or May \or June \or July \or August \or September
           \or October \or November \or December \fi}
\def\date{\hbox{\strut\mname \number\year}}
%
%
% Numbering of figures, tables and equations is automatic.
%
\newcount\FIGURENUMBER\FIGURENUMBER=0
\def\FIG#1{\expandafter\ifx\csname FG#1\endcsname\relax
               \global\advance\FIGURENUMBER by 1
               \expandafter\xdef\csname FG#1\endcsname
                              {\the\FIGURENUMBER}\fi}
\def\figtag#1{\expandafter\ifx\csname FG#1\endcsname\relax
               \global\advance\FIGURENUMBER by 1
               \expandafter\xdef\csname FG#1\endcsname
                              {\the\FIGURENUMBER}\fi
              \csname FG#1\endcsname\relax}
\def\fig#1{\expandafter\ifx\csname FG#1\endcsname\relax
               \global\advance\FIGURENUMBER by 1
               \expandafter\xdef\csname FG#1\endcsname
                      {\the\FIGURENUMBER}\fi
           Fig.~\csname FG#1\endcsname\relax}
\def\figand#1#2{\expandafter\ifx\csname FG#1\endcsname\relax
               \global\advance\FIGURENUMBER by 1
               \expandafter\xdef\csname FG#1\endcsname
                      {\the\FIGURENUMBER}\fi
           \expandafter\ifx\csname FG#2\endcsname\relax
               \global\advance\FIGURENUMBER by 1
               \expandafter\xdef\csname FG#2\endcsname
                      {\the\FIGURENUMBER}\fi
           figures \csname FG#1\endcsname\ and
                   \csname FG#2\endcsname\relax}
\def\figto#1#2{\expandafter\ifx\csname FG#1\endcsname\relax
               \global\advance\FIGURENUMBER by 1
               \expandafter\xdef\csname FG#1\endcsname
                      {\the\FIGURENUMBER}\fi
           \expandafter\ifx\csname FG#2\endcsname\relax
               \global\advance\FIGURENUMBER by 1
               \expandafter\xdef\csname FG#2\endcsname
                      {\the\FIGURENUMBER}\fi
           figures \csname FG#1\endcsname--\csname FG#2\endcsname\relax}
\newcount\TABLENUMBER\TABLENUMBER=0
\def\TABLE#1{\expandafter\ifx\csname TB#1\endcsname\relax
               \global\advance\TABLENUMBER by 1
               \expandafter\xdef\csname TB#1\endcsname
                          {\the\TABLENUMBER}\fi}
\def\tabletag#1{\expandafter\ifx\csname TB#1\endcsname\relax
               \global\advance\TABLENUMBER by 1
               \expandafter\xdef\csname TB#1\endcsname
                          {\the\TABLENUMBER}\fi
             \csname TB#1\endcsname\relax}
\def\table#1{\expandafter\ifx\csname TB#1\endcsname\relax
               \global\advance\TABLENUMBER by 1
               \expandafter\xdef\csname TB#1\endcsname{\the\TABLENUMBER}\fi
             Table \csname TB#1\endcsname\relax}
\def\tableand#1#2{\expandafter\ifx\csname TB#1\endcsname\relax
               \global\advance\TABLENUMBER by 1
               \expandafter\xdef\csname TB#1\endcsname{\the\TABLENUMBER}\fi
             \expandafter\ifx\csname TB#2\endcsname\relax
               \global\advance\TABLENUMBER by 1
               \expandafter\xdef\csname TB#2\endcsname{\the\TABLENUMBER}\fi
             Tables \csname TB#1\endcsname{} and
                    \csname TB#2\endcsname\relax}
\def\tableto#1#2{\expandafter\ifx\csname TB#1\endcsname\relax
               \global\advance\TABLENUMBER by 1
               \expandafter\xdef\csname TB#1\endcsname{\the\TABLENUMBER}\fi
             \expandafter\ifx\csname TB#2\endcsname\relax
               \global\advance\TABLENUMBER by 1
               \expandafter\xdef\csname TB#2\endcsname{\the\TABLENUMBER}\fi
            Tables \csname TB#1\endcsname--\csname TB#2\endcsname\relax}
\newcount\REFERENCENUMBER\REFERENCENUMBER=0
\def\REF#1{\expandafter\ifx\csname RF#1\endcsname\relax
               \global\advance\REFERENCENUMBER by 1
               \expandafter\xdef\csname RF#1\endcsname
                         {\the\REFERENCENUMBER}\fi}
\def\reftag#1{\expandafter\ifx\csname RF#1\endcsname\relax
               \global\advance\REFERENCENUMBER by 1
               \expandafter\xdef\csname RF#1\endcsname
                      {\the\REFERENCENUMBER}\fi
             \csname RF#1\endcsname\relax}
\def\ref#1{\expandafter\ifx\csname RF#1\endcsname\relax
               \global\advance\REFERENCENUMBER by 1
               \expandafter\xdef\csname RF#1\endcsname
                      {\the\REFERENCENUMBER}\fi
             [\csname RF#1\endcsname]\relax}
\def\refto#1#2{\expandafter\ifx\csname RF#1\endcsname\relax
               \global\advance\REFERENCENUMBER by 1
               \expandafter\xdef\csname RF#1\endcsname
                      {\the\REFERENCENUMBER}\fi
           \expandafter\ifx\csname RF#2\endcsname\relax
               \global\advance\REFERENCENUMBER by 1
               \expandafter\xdef\csname RF#2\endcsname
                      {\the\REFERENCENUMBER}\fi
             [\csname RF#1\endcsname--\csname RF#2\endcsname]\relax}
\def\refand#1#2{\expandafter\ifx\csname RF#1\endcsname\relax
               \global\advance\REFERENCENUMBER by 1
               \expandafter\xdef\csname RF#1\endcsname
                      {\the\REFERENCENUMBER}\fi
           \expandafter\ifx\csname RF#2\endcsname\relax
               \global\advance\REFERENCENUMBER by 1
               \expandafter\xdef\csname RF#2\endcsname
                      {\the\REFERENCENUMBER}\fi
            [\csname RF#1\endcsname,\csname RF#2\endcsname]\relax}
\def\refs#1#2{\expandafter\ifx\csname RF#1\endcsname\relax
               \global\advance\REFERENCENUMBER by 1
               \expandafter\xdef\csname RF#1\endcsname
                      {\the\REFERENCENUMBER}\fi
           \expandafter\ifx\csname RF#2\endcsname\relax
               \global\advance\REFERENCENUMBER by 1
               \expandafter\xdef\csname RF#2\endcsname
                      {\the\REFERENCENUMBER}\fi
            [\csname RF#1\endcsname,\csname RF#2\endcsname]\relax}
\def\Ref#1{\expandafter\ifx\csname RF#1\endcsname\relax
               \global\advance\REFERENCENUMBER by 1
               \expandafter\xdef\csname RF#1\endcsname
                      {\the\REFERENCENUMBER}\fi
             Ref.~\csname RF#1\endcsname\relax}
\def\Refto#1#2{\expandafter\ifx\csname RF#1\endcsname\relax
               \global\advance\REFERENCENUMBER by 1
               \expandafter\xdef\csname RF#1\endcsname
                      {\the\REFERENCENUMBER}\fi
           \expandafter\ifx\csname RF#2\endcsname\relax
               \global\advance\REFERENCENUMBER by 1
               \expandafter\xdef\csname RF#2\endcsname
                      {\the\REFERENCENUMBER}\fi
            Refs.~\csname RF#1\endcsname--\csname RF#2\endcsname]\relax}
\def\Refand#1#2{\expandafter\ifx\csname RF#1\endcsname\relax
               \global\advance\REFERENCENUMBER by 1
               \expandafter\xdef\csname RF#1\endcsname
                      {\the\REFERENCENUMBER}\fi
           \expandafter\ifx\csname RF#2\endcsname\relax
               \global\advance\REFERENCENUMBER by 1
               \expandafter\xdef\csname RF#2\endcsname
                      {\the\REFERENCENUMBER}\fi
        Refs.~\csname RF#1\endcsname{} and \csname RF#2\endcsname\relax}
\def\refadd#1{\expandafter\ifx\csname RF#1\endcsname\relax
               \global\advance\REFERENCENUMBER by 1
               \expandafter\xdef\csname RF#1\endcsname
                      {\the\REFERENCENUMBER}\fi \relax}

\newcount\EQUATIONNUMBER\EQUATIONNUMBER=0
\def\EQ#1{\expandafter\ifx\csname EQ#1\endcsname\relax
               \global\advance\EQUATIONNUMBER by 1
               \expandafter\xdef\csname EQ#1\endcsname
                          {\the\EQUATIONNUMBER}\fi}
\def\eqtag#1{\expandafter\ifx\csname EQ#1\endcsname\relax
               \global\advance\EQUATIONNUMBER by 1
               \expandafter\xdef\csname EQ#1\endcsname
                      {\the\EQUATIONNUMBER}\fi
            \csname EQ#1\endcsname\relax}
\def\EQNO#1{\expandafter\ifx\csname EQ#1\endcsname\relax
               \global\advance\EQUATIONNUMBER by 1
               \expandafter\xdef\csname EQ#1\endcsname
                      {\the\EQUATIONNUMBER}\fi
            \eqno(\csname EQ#1\endcsname)\relax}
\def\EQNM#1{\expandafter\ifx\csname EQ#1\endcsname\relax
               \global\advance\EQUATIONNUMBER by 1
               \expandafter\xdef\csname EQ#1\endcsname
                      {\the\EQUATIONNUMBER}\fi
            (\csname EQ#1\endcsname)\relax}
\def\eq#1{\expandafter\ifx\csname EQ#1\endcsname\relax
               \global\advance\EQUATIONNUMBER by 1
               \expandafter\xdef\csname EQ#1\endcsname
                      {\the\EQUATIONNUMBER}\fi
          Eq.~(\csname EQ#1\endcsname)\relax}
\def\eqand#1#2{\expandafter\ifx\csname EQ#1\endcsname\relax
               \global\advance\EQUATIONNUMBER by 1
               \expandafter\xdef\csname EQ#1\endcsname
                        {\the\EQUATIONNUMBER}\fi
          \expandafter\ifx\csname EQ#2\endcsname\relax
               \global\advance\EQUATIONNUMBER by 1
               \expandafter\xdef\csname EQ#2\endcsname
                      {\the\EQUATIONNUMBER}\fi
         Eqs.~\csname EQ#1\endcsname{} and \csname EQ#2\endcsname\relax}
\def\eqto#1#2{\expandafter\ifx\csname EQ#1\endcsname\relax
               \global\advance\EQUATIONNUMBER by 1
               \expandafter\xdef\csname EQ#1\endcsname
                      {\the\EQUATIONNUMBER}\fi
          \expandafter\ifx\csname EQ#2\endcsname\relax
               \global\advance\EQUATIONNUMBER by 1
               \expandafter\xdef\csname EQ#2\endcsname
                      {\the\EQUATIONNUMBER}\fi
          Eqs.~\csname EQ#1\endcsname--\csname EQ#2\endcsname\relax}
%
%THESE MACROS DEFINE SECTION AND SUBSECTION HEADERS
\newcount\SECTIONNUMBER\SECTIONNUMBER=0
\newcount\SUBSECTIONNUMBER\SUBSECTIONNUMBER=0
\def\section#1{\global\advance\SECTIONNUMBER by 1\SUBSECTIONNUMBER=0
      \bigskip\goodbreak\line{{\sectnfont \the\SECTIONNUMBER.\ #1}\hfil}
      \bigskip}
\def\subsection#1{\global\advance\SUBSECTIONNUMBER by 1
      \bigskip\goodbreak\line{{\subsectnfont
         \the\SECTIONNUMBER.\the\SUBSECTIONNUMBER.\ #1}\hfil}
      \smallskip}
%
%SOME SPECIAL SYMBOLS
\def\lsim{\raise0.3ex\hbox{$<$\kern-0.75em\raise-1.1ex\hbox{$\sim$}}}
\def\gsim{\raise0.3ex\hbox{$>$\kern-0.75em\raise-1.1ex\hbox{$\sim$}}}
%
%DEFINE JOURNAL NAMES

\def\NP{{\sl Nucl.\ Phys.\ }}
\def\PL{{\sl Phys.\ Lett.\ }}

\def\ZP{{\sl Z.\ Phys.\ }}

\def\manner{\hbox{\vbox{\offinterlineskip
                        \uctnum\binum\date}}\hfill\relax}
\footline={\ifnum\pageno=0\manner\else\hfil\number\pageno\hfil\fi}

\hsize=16.0 truecm
\vsize=23.0 truecm
\baselineskip=13 pt
\footline={\ifnum\pageno=0\hfil\else\hss\tenrm\folio\hss\fi}
\pageno=0
%name of file c:\jc\papers\texb.tex
%PAGE SIZE, FONTS, PREPRINT NUMBER ETC.
%
%
%
% Numbering of figures, tables and equations is automatic.
%
%
%
%
%

\def\uctnum{\hbox{UCT-TP 197/93\strut}}

\def\manner{\hbox{\vbox{\offinterlineskip
                        \uctnum\date}}\hfill\relax}
\footline={\ifnum\pageno=0\else\hfil\number\pageno\hfil\fi}
\hfill CERN-TH 6923/93 \par
\hfill BI-TP-93/32 \par
\vskip 3 truecm
\font\titlefont=cmbx10 scaled \magstep1
\begingroup\titlefont\obeylines
\centerline{TRANSITION FROM A QUARK-GLUON PLASMA}
\medskip
\centerline{IN THE PRESENCE OF A SHARP FRONT}
\endgroup\bigskip\bigskip
\parindent=0pt
\centerline{
N.~Bili\'c$^{1,2}$\footnote{}{1$~~\!$Department of
Physics, University of Cape Town, Rondebosch 7700, South Africa},
{}~J.~Cleymans$^{1}$\footnote{}{2$~~\!$Rudjer Bo\v{s}kovi\'c Institute,
P.O. Box 1016, Zagreb, Croatia},
{}~K.~Redlich$^{3,4}$\footnote{}{3$~~\!$Fakult\"at f\"ur Physik,
Universit\"at
Bielefeld, W-4800 Bielefeld, Germany}
and E.~Suhonen$^{5,6}$\footnote{}{4$~~\!$Department of Theoretical
Physics,
University of Wroclaw, PL-5025 Wroclaw, Poland}
\footnote{}{5$~\!$ Department of Theoretical Physics,
University of Oulu, SF-90550 Oulu, Finland}
\footnote{}{6$~\!$ CERN, CH-1211 Geneva 23, Switzerland}
}
\vskip 2truecm
\parindent=20pt
\bigskip\centerline{\bf ABSTRACT}\medskip
 The effect of a sharp front separating the quark-gluon plasma
 phase from the hadronic phase is investigated.
Energy-momentum conservation and baryon number conservation
 constrain the possible temperature jump across the front.
If one assumes
 that the temperature in the hadronic phase is $T\simeq$ 200 MeV ,
as has been suggested by numerous results from
relativistic ion collisions, one can
determine the corresponding temperature in the quark phase with the
help
of continuity equations across the front.
The calculations reveal
that the quark phase must be in a strongly supercooled state.
The stability of this solution with respect to minor
modifications is investigated. In particular
the effect of an admixture of hadronic matter in the quark phase
(e.g. in
the form of bubbles) is considered in detail.
In the absence of admixture the transition proceeds via a detonation
transition and is accompanied by a substantial super-cooling of
the quark-gluon plasma phase.
%
%**********************
%
The detonation is accompanied by less supercooling
if a small fraction of bubbles is allowed.
 By increasing the fraction of bubbles the supercooling becomes
 weaker and eventually the transition proceeds via a smoother
 deflagration wave.
%**********************
\refadd{NA35}
\refadd{WA85}
\refadd{NA36}
\refadd{NA44}
\refadd{satz1}
\refadd{satz2}
\refadd{rafelski1}
\refadd{rafelski2}
\refadd{rafelski3}
\refadd{pretoria1}
\refadd{pretoria2}
\refadd{pretoria3}
\refadd{heinz}
\refadd{karsch}
\vfill\eject
\section{Introduction}
Experimental results from heavy-ion collisions at CERN indicate that
hadrons are being produced with a temperature of about $T\simeq$  200
MeV
[\reftag{NA35},\reftag{WA85},\reftag{NA36},\reftag{NA44}].
{}From an analysis of the ratios of strange baryons as measured by the
WA85
collaboration at CERN [\reftag{WA85}]
one finds that a baryonic chemical potential of
$\mu_B$ = 300 MeV is consistent with their results
[\reftag{satz1}-\reftag{pretoria3}] (provided no transverse flow
exists
[\reftag{heinz}]).
As a first attempt one
can therefore assume that the thermodynamic properties of the
hadronic gas
are known. It is then of interest to investigate whether this
hadronic gas
could have been in a quark-gluon plasma phase before freeze-out. To
this
end we assume that a sharp front existed separating the two phases.
Various conservation laws , e.g. baryon number, energy and momentum,
restrict
the temperature difference across the front. In addition, the second
law
of thermodynamics requires the entropy to increase across the front.
It can easily be established that continuity in the various
conservation
laws leads to only one possibility, namely a highly supercooled
quark-gluon plasma must have existed
before the transition to the hadronic
phase. The degree of supercooling is surprisingly high, $T_q\sim$ 70
MeV.
This leads one to suspect whether some alternate mechanism started
operating, preventing the quark-gluon plasma from supercooling to
such a
large degree. In this paper we would like to investigate several
possibilities which could prevent this. In particular,
it is known  that supercooling enhances the
formation of bubbles of hadronic matter. We will therefore consider
that the quark phase was not pure but that some  admixture of
hadronic matter exists.
It is  known that the boiling of liquids
only takes place if impurities are present or if some
admixture of the gaseous state exists. Our results confirm this
standard
knowledge.
%
%+++++++++
 The transition from quark-gluon plasma to hadrons
proceeds in a violent manner towards superheated  hadrons.
The presence of an admixture of hadronic material in the form of
bubbles
or in any other form  makes the transition easier.
Our results agree with this expectation, from the moment the fraction
of hadronic material is  appreciable,
 the violent detonation type transition turns into a
smoother deflagration type. The transition between the two solution
shows
an S-type of curve. This means that for a given reaction the
solution is not unique.
%**************
%COMMENT:
%Because I hate to admit that we don't know how to interpret
%something , instead of:
%
% We do not know how to interpret this, it could be
%
%I would prefer something like
  A possible interpretation would be
that in this region a chaotic turbulent type of transition takes
place.

The paper is organized as follows : in the next section we discuss the
equations of state we use in the the quark-gluon plasma and in the
hadronic phases; in section 3 we present for completeness the
equations
expressing energy, momentum and baryon number  conservation. We also
discuss the adiabaticity condition; in section 4 we discuss the
admixture
of ``impurities'' into the quark phase in the form of hadronic
matter. The
velocities around the sharp front are discussed in section 5. In
section 6
we present our conclusions and give a summary of our results.
\section{Equation of State in the Quark Gluon Plasma}
In this section we discuss the equation of state used in the quark-
gluon
plasma phase. We consider
the cut-off model, motivated by numerical results
from lattice gauge theories [\reftag{karsch}]. The
precise value of the cut-off has a big influence on our results. For
comparison we also consider the M.I.T. bag equation of state
%*********************
which has been previously discussed in detail [\reftag{redlich}].
%********************
In the cut-off model one leaves out the low momentum part of the
particles, e.g. for gluons we have  :
$$
n_g=16\int_{k_c}^{\infty} {d^3p\over (2\pi )^3}{1\over e^{E/T}- 1}
\eqno(1)
$$

 It is well-known that the cut-off model
reproduces
the lattice results reasonably well. It is not excluded that a better
way
to proceed is to give the gluon a temperature dependent mass as has
been
argued recently by Goloviznin and Satz [\reftag{goloviznin}].
 This proposal cannot be followed
up here because we want to include the possibility of super-cooling
and
heating and we, therefore, have to know the equation of state both
below and above the critical temperature. We
have chosen a value of $5T_c$ for the cut-off
parameter in order to reproduce the critical temperature $T_c$ obtained in
lattice gauge theories for $\mu_B=0$ [\reftag{gottlieb}].
 Certain quantities are very sensitive to the precise value
of
this cut-off and to show this influence we also present for comparison
results obtained with a value of
$3.5T_c$.
%*********************
 The corresponding results for the energy density are shown
 in Fig. 1.
%*********************
On the hadronic side we incorporate the hard core radius of hadrons by
choosing [\reftag{helmut}]
$$
\epsilon_h={\epsilon_h^0\over 1+V_0n_h^0} ,
\eqno(2)
$$
where the subscript zero refers to point-like quantities, $V_0$
refers to
a typical proper volume of a hadron which we take to have a radius
given
by 0.8 $fm$.
The hadronic side in our approach is a composition of non-interacting
quantum gases. The use of quantum statistics is essential
since in a wide range
of $\mu_B$ and $T$ the Boltzmann approximation is not satisfied.
We include  all
well established hadronic resonances listed in
the latest edition of the
Review of Particle Properties [\reftag{Particles}]. A
correction similar to equation (11) is performed to all thermodynamic
quantities.
The behaviour suggested by the M.I.T.-bag equation of state
is very different and is not supported by the lattice results.
The larger cut-off reduces the energy-density in the quark-gluon
plasma
since a big contribution is being left out. The latent heat thus
becomes
very small for a large cut-off.
Since lattice results
are only available for zero chemical potential, $\mu_B$=0,
it is not immediately clear how to
generalize the cut-off model to
finite baryon densities.
%**********
Assuming the cut-off to be independent of $\mu_B$ we have
determined the phase diagram depicted in Fig. 2
where we compare it with the M.I.T.-bag model one.
%**********
 At $\mu_B=0$ the two are chosen
to coincide. For non-zero values of $\mu_B$ they become more and more
different. For the region we are interested in $\mu_B\sim 300$ MeV
and the
difference is not very large.
Having discussed the equations of state we will be using in each
phase, we
now proceed to the equations of continuity across the sharp front.
\section{Equations of Continuity across the Boundary}
Energy and momentum conservation across the front
leads to the
well-known two equations
(see e.g. [\reftag{Landau}]) :
$$
(\epsilon_h + P_h)\gamma_h^2 v_h + P_h =(\epsilon_q + P_q)\gamma_q^2
 v_q+P_q ,
\eqno(3)
$$
$$
(\epsilon_h+P_h)v_h^2\gamma_h^2=(\epsilon_q+P_q)v_q^2\gamma_q^2 ,
\eqno(4)
$$
both sides of the front are at  different temperatures and
chemical potentials.
An important remark here : $\epsilon_q$ refers to the quark-gluon
plasma
side of the sharp front, where we allow for an admixture of hadronic
matter
%**********
consisting for simplicity of only pions.
Thus
%**********
$$
\epsilon_q=(1-f)\epsilon_q^0+f\epsilon_{\pi}
\eqno(5)
$$
%*********
where $\epsilon_q^0$ refers to the pure quark-gluon matter
%*********
and $f$ is the fraction of hadronic matter present in the quark-gluon
plasma phase. We will consider $f$ to be a free parameter and
investigate how
various quantities depend on it. It turns out
that for the small cut-off as well as for the bag model the transition
temperature and chemical potential are not sensitive to variation of
$f$
provided that it is not very large.
Fore the large cut-off the transition changes completely its nature
when $f$ becomes of the order of 20\%.
%***************
The thermodynamic quantities (pressure, energy
density) refer to the rest frame quantities.
The index $q$ refers to the quark-gluon plasma side
which has an admixture of hadronic matter, while the index $h$
refers to the hadronic phase, $v$ refers to the velocity
of the gas in each phase with
respect to the front, $\gamma$ is the standard relativistic factor
($1/\sqrt{1-v^2})$.
In addition to the above two equations, baryon number conservation
across the front leads to
$$
n_hv_h\gamma_h=n_qv_q\gamma_q
\eqno(6)
$$
where $n_h$ and $n_q$ refer to the baryon densities
on the hadronic and on the quark-gluon plasma sides respectively.

In order to determine the four unknown quantities
$\epsilon_q, P_q, v_q$ and $v_h$ as functions of
$\epsilon_h, P_h$ and $n_h$ we need one more equation.
This is given by the requirement of non-decreasing entropy
which implies in our notation
$$
s_hv_h\gamma_h\geq s_qv_q\gamma_q \quad ,
\eqno(7)
$$
or, combining (6) and (7)
$$
{s_h\over n_h} \geq {s_q\over n_q} \quad .
\eqno(8)
$$
As has been shown previously [\reftag{redlich}], eqs. (3,4,6,7) yield
a
supercooled quark-gluon plasma.
The equality sign in (8), corresponding to an adiabatic process,
determines the minimal amount of supercooling needed to fit
the hadronic data without violating the second law of
thermodynamics. Therefore, we may use the equation
$$
{s_h\over n_h}={s_q\over n_q} ,
\eqno(9)
$$
as the fourth constraint, keeping in mind that the quark-gluon
temperature $T_q$ thus obtained is maximal.

Equations (3,4,6) can be solved for the velocities, leading to
$$
v_q^2={(P_h-P_q)(\epsilon_h+P_q)\over
(\epsilon_h-\epsilon_q)(\epsilon_q+P_h)},
\eqno(10)
$$
$$
v_h^2={(P_h-P_q)(\epsilon_q+P_h)\over
(\epsilon_h-\epsilon_q)(\epsilon_h+P_q)} .
\eqno(11)
$$
%++++++
%The change in density can then be determined from
%++++++++
%*************
With help of eq. (6) the velocities can be eliminated
yielding
%*************
$$
\left({n_q\over n_h} \right)^2=
{(\epsilon_q+P_q)(\epsilon_q+P_h)\over
(\epsilon_h+P_h)(\epsilon_h+P_q)}.
\eqno(12)
$$
%**************
Combining now equations (9) and (12) we can calculate numerically
the temperature $T_q$ and the chemical potential $\mu_B^q$.
This in turn yields all thermodynamic quantities on the
quark-gluon side.
 In the following sections we discuss our
main results.
%**************
%
%
%
\section{Admixture of Hadronic Matter}
In Fig. 3 we show our results for the case when there is no admixture
of
hadronic matter in the quark-gluon plasma phase. There is a
substantial
amount of supercooling in the quark phase. The transition then takes
place
to a hadronic gas which has a temperature $T= 200$ MeV as suggested by
most experiments at CERN using a relativistic ion beam. Most results
from
lattice gauge theories indicate that this temperature is higher than
the
critical temperature of the phase transition. We thus have to conclude
that the hadronic gas is in a super-heated phase.

This scenario immediately raises the following question : what
happens if
other mechanisms prevent this strong super-cooling? In particular it
has
been argued convincingly by Csernai and Kapusta
[\reftag{kapusta}] that super-cooling
enhances the formation of hadronic bubbles in the quark-gluon plasma
phase. We have therefore considered admixing hadronic matter to the
quark-gluon plasma phase.
We expected that the strong degree of supercooling would disappear
however, we did not expect our results to be so sensitive on the
value of
the cut-off chosen for the equation of state in the quark-gluon plasma
phase.
This is shown in Fig. 4. In this figure we keep the parameters on the
hadronic side fixed at $T=200$ MeV and $\mu_B=300$ MeV but we vary the
amount of hadronic admixture, given by the fraction $f$ as
introduced in equation (4), on the quark side and we calculate each
time
the resulting value of the temperature on the other side of the sharp
front.
For simplicity the
admixture is in the form of pions only. If the cut-off value is large,
$5T_c$ , then there is an abrupt turnover when the admixture exceeds
20\%.
The temperature increase across the front is reduced after this.
However
when the cut-off is taken at $3.4T_c$ then the picture is completely
different : the strong supercooling persists anomalously long until
the
admixture has reached more than 60\%, in which case it is no longer
possible to speak of a ``small'' admixture.
%********
If a bag-model equation of state is used the behavior is very
similar to the case of the small cut-off.
%********

In Fig. 5 we show the baryon chemical potential
$\mu_q$ on the quark-gluon plasma side. The observed behaviour
follows the
pattern already found for the temperature in Fig. 4.
As previously, the cut-off in the
equation of state is very important.

\section{Velocities}
\refadd{Vanhove}
\refadd{allbigshots}
It is customary to work in the frame where the sharp front separating
the
two phases is at rest.
In this frame the quark-gluon plasma moves towards the front with a
velocity $v_q$ and the hadrons move away from the front with a
velocity
$v_h$.  If the two velocities are different, as is usually the case,
one
observes a phenomenon of pile up on one side, e.g. if the quarks have
the
largest velocity, the hadrons will pile up with a large baryon density
at the other side of the front.
If $v_q<v_h$ one has a deflagration, if $v_q>v_h$ one has a detonation
(see [\reftag{Vanhove}-\reftag{russians}]).

The velocities can be easily calculated from equations (10) and (11).
Our results for
$v_q^2$ and $v_h^2$ as functions of the fraction $f$ of hadronic matter
mixed  into the quark-gluon phase are shown in Figs. 6 and 7 respectively.
In Fig. 8 we plot the ratio of $v_h$ to $v_q$ as a function of $f$ and in
Fig. 9 the relative velocity defined as
$$
v_{rel}={v_h-v_q\over 1-v_hv_q}
\eqno(13)
$$
The change
in the transition mechanism leads to very sharp variations in $v_q$
and $v_h$. The velocities  become unphysical along the curves between the
points A and A'. The detonation and deflagration branches are clearly separated
in Figs. 6 and 7 whereas in Figs. 8 and 9 the detonation takes place below the
point A and deflagration above the point A'.

As is well known, small perturbations in a medium  propagate with the
velocity of sound in that medium. It is therefore of special interest
to
compare $v_q$ and $v_h$ with the velocity of sound in each phase.
The latter is defined as [\reftag{Landau}]
$$
c_s^2=\left.{\partial P\over\partial\epsilon}\right|_{s/n} .
\eqno(14)
$$
 To calculate the speed of sound
we write eq. (14) as
$$
c_{s}^2={\partial P/\partial T +\partial P/\partial\mu_B\cdot
d\mu_B/dT
      \over
 \partial\epsilon/\partial T +\partial\epsilon/\partial\mu_B\cdot
d\mu_B/dT} ,
\eqno(15)
$$
where the  derivative
$d\mu_B/dT$ is taken such that the ratio $s_h/n_h$ remains constant.
Thus
$$
{d\mu_B\over dT}=-{n\partial s/\partial T -s\partial
n/\partial T \over n\partial s/\partial\mu_B -s\partial
n/\partial\mu_B} .
\eqno(16)
$$

The velocities of sound in the quark-gluon  and in the hadronic phases are
shown in Figs. 6 and 7 as functions of $f$.

The points O, O', A and A' divide the curves in Figs. 6-9 according to
the magnitudes of the velocities of sound in the corresponding media. The
following inequalities hold :\par
below O  : $v_q>c_{s,q} \quad ,\quad v_h<c_{s,h}$\par
on OA    : $v_q<c_{s,q} \quad ,\quad v_h<c_{s,h}$\par
on A'O'  : $v_q>c_{s,q} \quad ,\quad v_h>c_{s,h}$\par
above O' : $v_q<c_{s,q} \quad ,\quad v_h>c_{s,h}$\par

The comparison of $v_q$ and $v_h$ with the velocities of sound in the
corresponding media shows that the conditions
for the stability of the front
$$
v_q>c_{s,q}\quad ;\quad v_h<c_{s,h}
\eqno(17)
$$
are
 fulfilled only along the upper part of the detonation branch. Following the
arguments presented in Ref. [\reftag{Landau}] one would conclude that the rest
of the curve is unstable and the front would disintegrate. Condition (17) is
however derived under the assumption of a
non-decreasing entropy density, i.e.
for $s_h\geq s_q$. In the present situation we require only the entropy {\sl
per baryon} not to decrease which means that $s_h<s_q$ is physically
acceptable provided the condition expressed in eq. (8) is satisfied.
Indeed we find that the ratio $s_q/s_h$ becomes less than one along the
detonation branch (corresponding to the part below A' on Figs. 8 and 9)
and $s_q/s_h>1$ along the deflagration branch (above A').
\section{Conclusions}
In this paper we have investigated the consequences arising from the
existence of a sharp front separating the quark-gluon plasma phase
from
the hadronic phase. The equations of continuity for energy, momentum
and
baryon number as well as the constraint following from the second law
of
thermodynamics (entropy increase), require  a substantial amount
of
supercooling to occur in the quark-gluon plasma before the transition
can take place. This transition then leads to a highly super-heated
hadronic gas. We have also investigated what happens if the quark-
gluon
plasma is not pure but contains an admixture of hadronic matter. It is
known that supercooling is limited when ``impurities'' are present in
the
super-cooled phase. We found that from the  moment when the admixture
of
hadronic matter is more than about 20\% of the total (the fraction
refers
to the energy content, not to the particle content) then the
transition
does not take place via a detonation type of wave but via a
deflagration
one. In this case the transition is  smoother and the temperature jump
across the sharp front is not as pronounced as in the absence of
admixture of hadronic matter.

%***************
 The crossover region between the two solution shows a remarkable
 S-shape structure. The multiple valued $T_q$ and $\mu_B^q$
 in that region indicate that in the given range of $f$
 a chaotic turbulent type of transition could take place.
%****************
%
%
%
\bigskip
In summary, we have considered a model based on the existence of a
sharp
front separating the hadronic from the quark-gluon phase.
We have arrived at the conclusion
that in this situation the transition can only proceed if a
substantial
amount of super-cooling takes place in the quark-gluon plasma phase
if the transition is between pure phases.
An admixture of hadronic material in the quark-gluon plasma phase
makes
the transition proceed in a much smoother way.
\bigskip
\bigskip
\bigskip
\noindent
{\bf Acknowledgment} We acknowledge
continuing stimulating discussions in this field
with H. Satz.
Two of us (E.S. and K.R.) thank the Physics
Department of the University of Cape Town for its hospitality.
N.B. and J.C. would like to thank the hospitality of the
Fakult\"at f\"ur Physik of the University of
Bielefeld where this work was completed. (KR) also acknowledges
the financial support by the
German Federal Ministry for Science and Technology (BMFT).
\vfill\eject
\noindent{\bf References}
\medskip
\item{\reftag{NA35})}
{J. Bartke et al. (NA35), \ZP C48 (1990) 191;\par
R. Stock et al. (NA35), \NP A525 (1991) 221c;\par
\vskip 3pt
\item{\reftag{WA85})}
{S. Abatzis et al. (WA85), \PL B244 (1990) 130;\par
S. Abatzis et al. (WA85), \PL B259 (1991) 508.}
\vskip 3pt
\item{\reftag{NA36})}
E. Andersen et al., Phys. Lett. B294 (1992) 127.
\vskip 3pt
\item{\reftag{NA44})}
{A. Franz (NA44), Report at the XXVI
International Conference on High Energy Physics, 6.-12.8.1992,
Dallas/Texas, USA.}
\vskip 3pt
\item{\reftag{satz1})}
J. Cleymans and H. Satz, Z. Phys., C57 (91993) 135.
\vskip 3pt
\item{\reftag{satz2})}
J. Cleymans, K. Redlich, H. Satz and E. Suhonen, Z. Phys., C58 (1993) 347.
\vskip 3pt
\item{\reftag{rafelski1})}J. Rafelski, Phys. Lett. B262 (1991) 333.
\vskip 3pt

\item{\reftag{rafelski2})}J. Rafelski, Nucl. Phys. A544 (1992) 279c.
\vskip 3pt
\item{\reftag{rafelski3})}
{J. Letessier,  A. Tounsi, U. Heinz, J. Sollfrank and J. Rafelski,
Paris Preprint PAR/LPTHE/92-37, September 1992.}
\vskip 3pt
\item{\reftag{pretoria1})}N.J. Davidson, H.G. Miller, R.M. Quick and
J.
Cleymans, Phys. Lett. B255 (1991) 195.
\vskip 3pt
\item{\reftag{pretoria2})}D.W. von Oertzen, N.J. Davidson, R.A.
Ritchie
and H.G. Miller, Phys. Lett. B274 (1992) 128.
\vskip 3pt
\item{\reftag{pretoria3})}N.J. Davidson, H.G. Miller, D.W. von Oertzen
 and K. Redlich, Z. Phys. C56 (1992) 319.
\vskip 3pt
\item{\reftag{heinz})} E. Schnedermann and U. Heinz,
 Phys. Rev. Lett. 69 (1992) 2908.
\vskip 3pt
\item{\reftag{karsch})} F. Karsch, Z.  Phys. C38 (1988) 147;\par
    M. Gorenstein and O. Mogilevsky, Z. Phys. C38 (1988)
161;\par
J. Engels, J. Fingberg, K. Redlich, H. Satz and M. Weber,\par
 Z.  Phys. C42 (1989) 341;\par
J. Engels, J. Fingberg, F. Karsch, D. Miller and M. Weber, \par
Phys.  Lett. B252 (1990) 625.
\vskip 3pt
%
%
%*****************
\item{\reftag{redlich})} N. Bili\'c, J. Cleymans, E.
Suhonen and D.W. von Oertzen, Cape Town preprint UCT-TP 93/191,
 Phys. Lett. B (1993) (in press).
\vskip 3pt
%*****************
%
\item{\reftag{goloviznin})} V.V. Goloviznin and  H. Satz,
   CERN preprint CERN-TH /93.
\vskip 3pt
\item{\reftag{gottlieb})} S. Gottlieb et al., Phys. Rev. Lett. 59 (1987)
1881.
\vskip 3pt
\item{\reftag{helmut})} J. Cleymans, K. Redlich, H. Satz and E.
Suhonen,
   Z. Phys. C33 (1986) 151.
\vskip 3pt
\item{\reftag{Particles})} Particle Data Group, Phys. Rev. D45 (1992)
S1.
\vskip 3pt
\item{\reftag{Landau})} L. Lan\-dau and E.M. Lif\-shitz,
    Hy\-dro\-dy\-na\-mics, Per\-ga\-mon Press, London (1959).
\vskip 3pt
\item{\reftag{kapusta})} L. Csernai and J. Kapusta,
    Phys. Rev. Lett. 67  (1993) 737.
\vskip 3pt
\item{\reftag{Vanhove})}L. Van Hove, Z. Phys. C21 (1983)
93; C27 (1985) 135.
\vskip 3pt
\item{\reftag{allbigshots})}M. Gyulassy, K. Kajantie, H. Kurki-Suonio
 and
  L. McLerran, Nucl. Phys. B237 (1984) 477.
\vskip 3pt
\item{\reftag{thorne})}K.S. Thorne, Astrophys. J. 179 (1973) 897.
\vskip 3pt

\vfil\eject
\noindent {\bf Figure Captions:}
\medskip
\item{1:}
Energy-density as a function of temperature at fixed
$\mu_B = 0.3$ 0.3 GeV.
The long-dashed line corresponds to the
equation of state of the M.I.T.-bag
with bag constant B=0.2 GeV/fm$^3$, the dashed
line is the cut-off model with a value $3.4T_c$ for the cut-off while
for
the long dash-dotted line the cut-off is $5T_c$. The solid line
corresponds to the energy-density as calculated from the hadronic gas.
The critical temperature is
indicated by the dotted line at $T_c=150$ MeV.
\vskip 3pt
\item{2:}The phase diagram calculated in the cut-off model and in the
M.I.T. bag model. At $\mu_B=0$ the bag constant is chosen such as to
reproduce the same critical temperature as the cut-off model. The
value of
the cut-off is taken to be $5T_c$ independent of $\mu_B$.
\vskip 3pt
\item{3:}The energy-density of the hadronic gas (lower curve)
and of the quark-gluon plasma (upper curve) as a function
of temperature. Super-cooling and super-heating are indicated by the
dashed line.
The transition
from the super-cooled quark-gluon plasma phase to the hadronic gas
phase
is indicated by the long-dashed line.
\vskip 3pt
\item{4:}The quark temperature $T_q$ versus the
fraction of hadronic admixture in the quark-gluon plasma phase.
The solid line corresponds to a cut-off value of $5T_c$, the
dashed line to $3.4T_c$. The dash-dotted line is for the equation of
state
of the M.I.T.-bag model.
\vskip 3pt
\item{5:}The baryon chemical potential in the quark-gluon plasma phase
versus the
fraction of hadronic admixture in the quark-gluon plasma phase.
The solid line corresponds to a cut-off value of $5T_c$,  the
dashed line to $3.4T_c$. The dash-dotted line is for the equation of
state
of the M.I.T.-bag model.
\vskip 3pt
\item{6:}The velocity squared of the quark matter (full line).
The value of the
cut-off parameter in the equation of state is $5T_c$, $ f$ is the
admixture
fraction. The dashed line is the velocity of sound squared in quark
matter.
See text for further explanations.
\vskip 3pt
\item{7:}The velocity squared of the  hadronic gas (full line).
The value of the
cut-off parameter in the equation of state is $5T_c$, $ f$ is the
admixture
fraction. The dashed line is the velocity of sound in the hadronic
matter.
See text for further explanations.
\vskip 3pt
\item{8:}Ratio of  the hadronic gas velocity
over the quark phase velocity as
a function of the admixture fraction f.
The value of the
cut-off parameter in the equation of state is $5T_c$.
If the ratio is less than one,
the transition proceeds via a detonation wave, if it is above one, it
proceeds via a deflagration wave. Between the points A and A' the velocities
are unphysical.
See text for further explanations.
\item{9:}The difference between the two velocities.
The value of the
cut-off parameter in the equation of state is $5T_c$, $f$ is the
admixture  fraction.  Between the points A and A' the velocities are unphysical
See text for further explanations.
\vfil
\eject
\bye
\end